# Calculation of Carrier Doping-Induced Half-Metallicity, and Transformation of Easy Axis in Two-Dimensional $MSi_2N_4$ (M = Cr, Mn, Fe, and Co) Monolayers

Ziyuan An,[†] Linhui Lv,[†] Ya Su,[ξ*] Yanyan Jiang,[‡] Zhaoyong Guan[†§*]

[†]Key Laboratory of Colloid and Interface Chemistry, Ministry of Education, School of Chemistry and Chemical Engineering, Shandong University, Jinan, Shandong 250100, P. R. China

[ξ]School of Electrical Engineering, Shandong University, Jinan, Shandong 250100, P. R. China

[‡]Key Laboratory for Liquid-Solid Structural Evolution & Processing of Materials (Ministry of Education), School of Materials Science and Engineering, Shandong University, Jinan, Shandong, 250061, People's Republic of China

[§]School of Chemistry and Chemical Engineering, Shandong University, Jinan 250100, P. R. China

## ABSTRACT

We study the stability, electrical properties, and magnetic properties of $MSi_2N_4$ (M = Cr, Mn, Fe, and Co) monolayers (MLs) based on the density functional theory (DFT). The calculation results show that the $MSi_2N_4$ MLs present the intrinsic antiferromagnetic (AFM) orders resulting from direct exchange interaction between the transition metal atoms. $CrSi_2N_4$ and $CoSi_2N_4$ MLs are half-metal (HM). Their spin-β gaps are 3.661 and 2.021 eV, respectively. $MnSi_2N_4$ and $FeSi_2N_4$ MLs are semiconductors with gaps of 0.427 and 0.282 eV, respectively. $MSi_2N_4$ MLs are stable in dynamics and thermodynamics. The effects of carrier doping on the electromagnetic properties of $MSi_2N_4$ MLs are also researched. We find that charge doping could transform $CrSi_2N_4$, $MnSi_2N_4$ and $CoSi_2N_4$ MLs from AFM to the ferromagnetic (FM) orders. We also find that $CrSi_2N_4$ and $CoSi_2N_4$ MLs are transformed from spin-polarized metal to HM by carrier doping. Moreover, the direction of doped $CrSi_2N_4$ ML's easy magnetization axis (EA) has been switched to the [001] axis when +0.9 *e* or more holes are doped. $MSi_2N_4$ family show an intrinsic antiferromagnetic order, good thermodynamic and kinetic stability, tunable magnetism and half-metallicity, suggesting their great potential applications in the spintronics.

## 1. INTRODUCTION

Since the graphene[1] is successfully prepared by mechanical stripping as the first two-dimensional (2D) material, 2D material have attracted much attention due to their excellent electrical,[2] thermal,[2] mechanical,[3, 4] chemical[5] catalytic performance. 2D material have an exciting prospect.[6-8] However, for a long time, the long-range magnetic order in 2D material has been regarded as non-existing at finite temperatures due to the existence of thermal fluctuations based on Mermin-Wagner theorem.[9] However, it is exciting that single atomic layered 2D ferromagnetic (FM) $CrI_3$ has been synthesized.[10] After $CrI_3$, other layered 2D ferromagnetic materials[11-13] like $Fe_3GeTe_2$,[14] $Cr_2Ge_2Te_6$,[15, 16] $2H-VSe_2$[17] were successfully synthesized and characterized. These ferromagnetic materials are expected to promote development of spintronics.[18-20]

Layered $MoSi_2N_4$[21] is a special 2D monolayer. Compared with other 2D materials,[22-27] there isn't corresponding three-dimensional parent material. Therefore, to study the layered $MoSi_2N_4$ is significant for us to broaden the sources and applications of 2D materials. However, the layered $MoSi_2N_4$ is paramagnetic,[28] which inhibits its wide prospect in the spintronics.[29, 30] Based on the above fact, 2D FM material families need to be enriched urgently,[31] however, to find intrinsic and controllable 2D FM materials is challenging.[32] At present, the commonly used methods to change 2D materials include charge doping,[33] defect engineering,[34] stress engineering,[35] magnetic field,[36] intercalation,[37, 38] and optical controlling,[39] but these methods are hard to implement in the experiment.[40] Therefore, it is of great significance[41] to design and research 2D FM materials with excellent performance. Gratifyingly, we can accomplish it by adopting computer technology to reduce the high cost of the experiments. Charge doping is a feasible and effective method[42] to control electromagnetic properties of 2D materials.

This paper focuses on the $MSi_2N_4$ MLs with outstanding performance. Based on the first-principles calculation method, systematic research on $MSi_2N_4$ is carried by DFT. In this work, $MSi_2N_4$ MLs' stability, electronic structure, and magnetism are analyzed. Moreover, the origin of the AFM order is revealed. The calculated results reveal that $MSi_2N_4$ MLs show AFM orders. However, $CrSi_2N_4$ and $CoSi_2N_4$ MLs with FM orders are HM. The EA of $CrSi_2N_4$ is switched from [100] to [001] directions under charge doping ($>+0.9\,e$). Moreover, the transition of magnetic order in $CrSi_2N_4$ and $MnSi_2N_4$ is realized by charge doping. In a word, this study accelerates the research process of 2D magnetic materials and promotes their wide applications in the field of spintronics.

## 2. COMPUTATIONAL DETAILS

The geometries of $MSi_2N_4$ are searched by adopting the particle-swarm optimization (PSO) method with the CALPSO code and ASAP.[43] The plane-wave basis Vienna ab initio simulation package (VASP) code[44] is used to calculate $MSi_2N_4$ MLs. Perdew−Burke−Ernzerhof (PBE)[45] is adopted to dealt with 3d-electrons. The hybrid Heyd−Scuseria−Ernzerhof

(HSE06) [46, 47] functional and GGA + U method[48] is applied to deal with the strong-correlated correction to the 3d electrons, respectively. The effective onsite Coulomb interaction parameters ($U$) are set to be 4.6 eV ($CrSi_2N_4$ and $MnSi_2N_4$), 6.0 eV ($FeSi_2N_4$), 7.6 eV ($CoSi_2N_4$). The exchange interaction parameters ($J_0$) are set to be 0.6 eV ($CrSi_2N_4$, $MnSi_2N_4$ and $CoSi_2N_4$), 0.5 eV ($FeSi_2N_4$), respectively. Therefore, the effective $U_{eff}$ ($U_{eff} = U - J_0$) are set as 4.0 eV ($CrSi_2N_4$ and $MnSi_2N_4$), 5.5 eV ($FeSi_2N_4$), 7.0 eV ($CoSi_2N_4$), which are used in the calculation of the magnetic anisotropy energy (MAE), phonon spectra, and ab initio molecular dynamics (AIMD) simulation. Lattice constant c along z-direction is set as 23 Å. The distance between adjacent interlayer atoms is 17 Å. The kinetic energy cutoff is set as 520 eV for all $MSi_2N_4$ MLs. The criteria of energy and force are is $10^{-6}$ eV and 10 meVÅ$^{-1}$, respectively. The geometry optimization and energy computation are adopted 3×3×1 and 9×9×1 Monkhorst−Pack $k$-grids, respectively. After the calculation of self-consistent is finished, total energies are evaluated by the noncollinear nonself-consistent calculations. The MAE is calculated with an energy cutoff of 600 eV, and the corresponding energy criterion of $1 \times 10^{-8}$ eV. The corresponding k-grid is adopted 20×20×1 with no symmetry constriction. Based on the finite displacement method, the phonon spectra and the phonon density of the states (PHDOS) are calculated in Phonopy package.[49] To calculate phonon spectra and the PHDOS, a 4×4×1 supercell is adopted. The corresponding criteria of energy is set as $10^{-8}$ eV and the Hellmann−Feynman force is 1 meVÅ$^{-1}$, respectively. 6,600 uniform k-points, which are along high-symmetry lines are adopted to get phonon spectra. AIMD simulation is used to prove geometry stability. The constant moles−volume−temperature ensemble with a Nosé−Hoover thermostat[50] is used at temperature of 300 K. The time step is 1 fs and the total time is 10 ps. To eliminate the effect of periodic boundary condition in smaller system size, a 2×2×1 cell (supercell) is adopted in AIMD simulation.

## 3. RESULTS AND DISCUSSION

### 3.1. Geometries of MSi$_2$N$_4$ MLs.

The geometries of MSi$_2$N$_4$ MLs are calculated and confirmed by PSO[43] and ASAP based on the crystal structure analysis. The optimal geometry of unit cell of MSi$_2$N$_4$ MLs is shown in the Fig S1, and the optimal geometries of supercells of MSi$_2$N$_4$ MLs are shown in Fig 1a-c. The lattice constants of MSi$_2$N$_4$ MLs are 2.909 (M = Cr), 2.920 (Mn), 2.920 (Fe) and 2.875 (Co) Å as presented in Table 1, respectively. MoSi$_2$N$_4$ ML's lattice constant is 2.910 Å.[28] The atomic radii of M and Mo are 118 (Cr), 117 (Mn), to 117 (Fe), 116 (Co) and 130 pm (Mo), which are close to each other. As a result, the corresponding lattice parameters of MSi$_2$N$_4$ are similar. The distance between M and nitrogen atoms ($d_{M-N}$) is 2.049, 2.052, 2.052, and 2.048 Å, respectively.[51] The $d_{M-N}$ of M is close to each other because of the similar atomic radii of M. More information about MSi$_2$N$_4$ could be found in Table 1. The

optimized

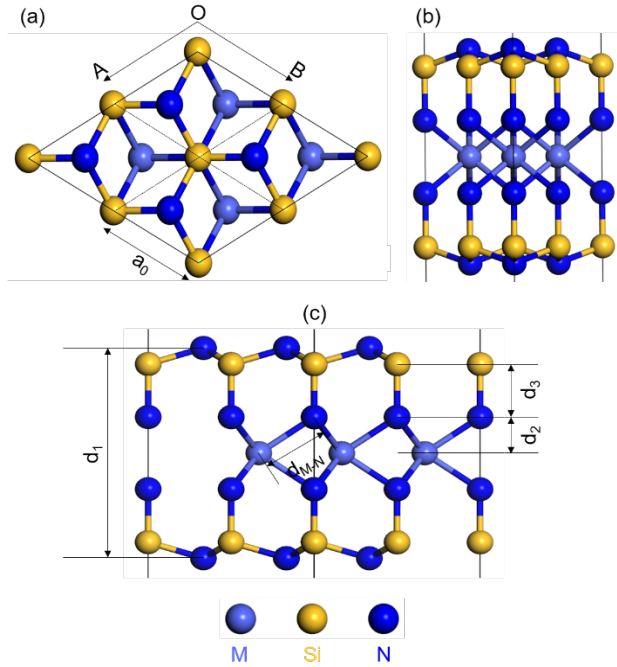

**Fig 1.** Optimized geometries of (a) top and (b, c) side views of $MSi_2N_4$ MLs. The left, medium, and right balls at the bottom of Fig 1c represent M, Si, and N atoms, respectively.

geometries of $MSi_2N_4$ MLs are shown in Fig 1a-c. The $d_1$, $d_2$, $d_3$, and $d_{M-N}$ in Table 1 are marked in Fig 1c. For example, the $d_1$ presents the distance from the top N atom to the bottom N atom.

**Table 1.** Calculated Lattice Constants $a_0$, Total Energies $E_t$, and Several Interatomic Distances for the AFM Orders of $MSi_2N_4$

| System | $a_0$(Å) | $d_1$(Å) | $d_2$(Å) | $d_3$(Å) | $d_{M-N}$(Å) |
| --- | --- | --- | --- | --- | --- |
| $CrSi_2N_4$ | 2.909 | 6.879 | 1,174 | 1.750 | 2.049 |
| $MnSi_2N_4$ | 2.920 | 6.858 | 1.170 | 1.745 | 2.052 |
| $FeSi_2N_4$ | 2.920 | 6.858 | 1.170 | 1.745 | 2.052 |
| CoSi$_2$N$_4$ | 2.875 | 6.989 | 1.199 | 1.762 | 2.048 |

**3.2. Electronic and Magnetic Properties of Unit Cell.** A single unit cell of MSi$_2$N$_4$ consists of a transition metal, two Si and four N atoms. There are electrons transfers from M atom to the nearby Si and N atoms due to electronegativity difference. Besides, the electron transfers make some d orbitals are occupied by just a single electron. Their spin charge densities are shown in Fig S2a-d. Each M atom has a magnetic moment (MM) of 2.56 (M = Cr), 3.31 (Mn), 3.84 (Fe), and 2.51 (Co) $\mu_B$, respectively, while the MMs of silicon and nitrogen atoms are closed to 0 $\mu_B$. The band structure and partial density of state (PDOS) of MSi$_2$N$_4$ are calculated, as shown in Fig 2a-d.

M atom is bonded with silicon and nitrogen atoms. Due to the electronegativity difference between atoms, there are charge transfers. The valence electron numbers of a M atom are 6.00 (Cr), 7.00 (Mn), 8.00 (Fe), and 9.00 (Co), respectively. The Bader charge analysis find that the numbers of valence electrons per atom are 4.48 (Cr), 5.44 (Mn), 6.50 (Fe), and 7.87 (Co) in MSi$_2$N$_4$ MLs, respectively. These valence electrons mainly fill the 3d orbitals. According to the integrated density of states (IDOS), the numbers of spin-β valence electrons of the MSi$_2$N$_4$ ML are 0.92 (Cr), 0.79 (Mn), 1.16 (Fe), and 2.12 (Co), respectively, as shown in Fig S3a-d. The estimated values of M/MMs of the MSi$_2$N$_4$ MLs are equal to the difference between the numbers of spin-α and spin-β valence electrons. These numbers can't exceed 5 as the limited number of 3d orbitals. So, based on the above results, it can be estimated that the M/MMs are 2.64 (Cr), 2.86 (Mn), 3.84 (Fe), and 2.88 (Co) $\mu_B$, respectively. They are close to the M/MMs calculated with HSE approaches, especially for $CrSi_2N_4$, $FeSi_2N_4$, and $CoSi_2N_4$ MLs. Given the Bader charge analysis and the IDOS, the charge transfers during the bonding process could well explain the M/MMs calculated with HSE approaches.

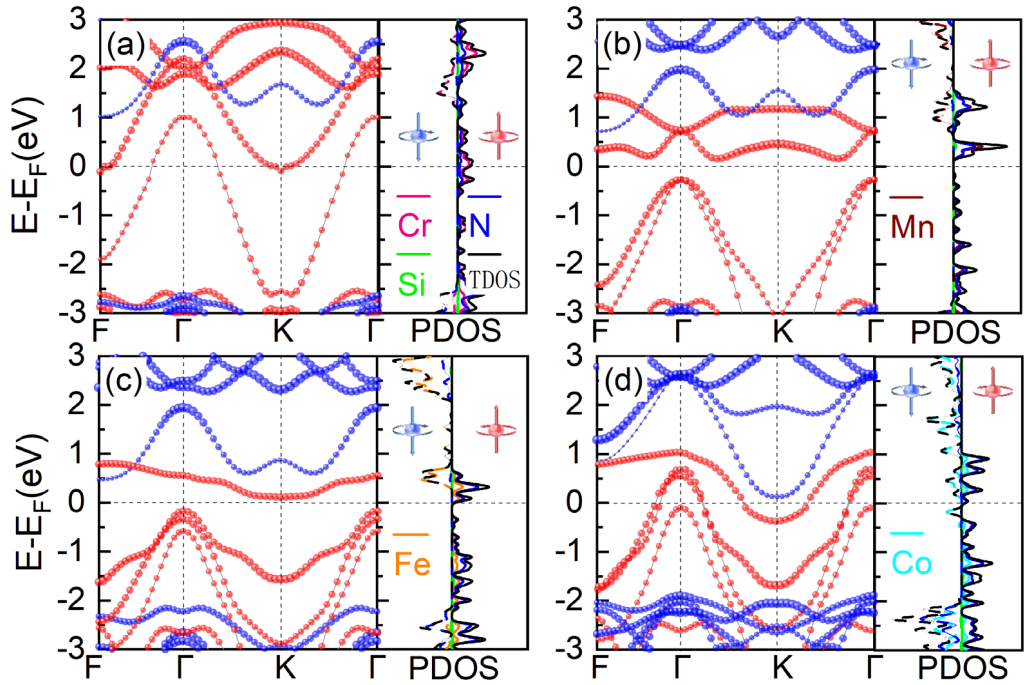

**Fig 2.** Band structures of FM orders of $MSi_2N_4$ MLs with (a) $CrSi_2N_4$, (b) $MnSi_2N_4$, (c) $FeSi_2N_4$, and (d) $CoSi_2N_4$ are calculated with HSE06 functional.

Besides magnetic properties, the electronic properties of $MSi_2N_4$ MLs are also researched. For $CrSi_2N_4$ and $CoSi_2N_4$, some bands occupied by spin-α electrons cross the Fermi energy, making them metallic, while the spin-β electrons of $CrSi_2N_4$ and $CoSi_2N_4$ MLs behave like semiconductive properties. Therefore, $CrSi_2N_4$ and $CoSi_2N_4$ MLs are HM. The Fermi-level is partially occupied by the spin-α electrons. For spin-β electrons in $CrSi_2N_4$ and $CoSi_2N_4$ MLs, the valance band maximum (VBM) of $CrSi_2N_4$ and $CoSi_2N_4$ is located at the Γ point, while the conduction band minimum (CBM) is located at the F and K points, respectively. There are about a bandgap of 3.661 eV and 2.021 eV in the $CrSi_2N_4$ and $CoSi_2N_4$ MLs' spin-β state as shown in Fig 2a,d, respectively. HM could provide 100% spin-polarized current, so the $CrSi_2N_4$ and $CoSi_2N_4$ MLs could work as the HMs. It means that they have a wide application in the spin transport device and spin injection.[52, 53] The p-orbital and d-orbital projected band structures of $CrSi_2N_4$ and $CoSi_2N_4$ MLs are calculated, shown in Fig 3a-d. Taking $CrSi_2N_4$ ML as an example, the states near the Fermi-level are contributed by N's $p_y$ orbitals, Cr's $d_{xy}$ and $d_{x^2-y^2}$ orbitals which are occupied by the spin-α electrons, as shown in Fig 3a, S4a-c, and S5a, respectively.

For $MnSi_2N_4$ and $FeSi_2N_4$ MLs, they are semiconductors, and the VBMs are located at the Γ point, as shown in Fig 2b,c, respectively. The CBM of $MnSi_2N_4$ is located at the point between the K and Γ points, and the CBM of $FeSi_2N_4$ is located at the K point. The calculated band gaps of $MnSi_2N_4$ and $FeSi_2N_4$ are 0.427 and 0.282 eV, respectively. The band structures indicate that the $MnSi_2N_4$ and $FeSi_2N_4$ MLs' orbitals near the Fermi-level are occupied by the spin-α electrons, shown in Fig 2b,c, respectively. The orbital projected band structures of MnSi$_2$N$_4$ and FeSi$_2$N$_4$ MLs are calculated, shown in Fig 2a-d. Taking FeSi$_2$N$_4$ ML as an example, the states near the Fermi-level are mainly contributed by N's p orbitals, while the states around the Fermi-level are partially contributed by the Fe's $d_{yz}$ and $d_{xz}$ orbitals, as shown in Fig 3c, S5c, and S6, respectively. Moreover, these orbitals are all occupied by the spin-α electrons. For other MLs, the p, $d_{xy}$, $d_{yz}$, and $d_{xz}$ orbitals make the contribution to the states near the Fermi-level mainly as shown in Fig 3.

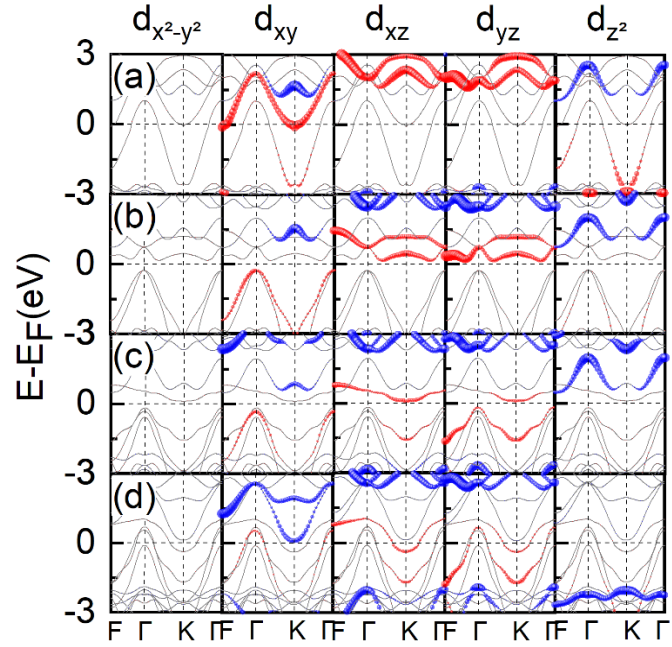

**Fig 3.** MSi$_2$N$_4$ MLs' d-orbital projected band structures. The red and blue lines with dots in every thumbnail image present $d_{x^2-y^2}$, $d_{xy}$, $d_{xz}$, $d_{yz}$, and $d_{z^2}$ atomic orbitals of (a) CrSi$_2$N$_4$, (b) MnSi$_2$N$_4$, (c) FeSi$_2$N$_4$, (d) CoSi$_2$N$_4$ from left to right. The Fermi-level is set 0.

After discussing the band structures without spin-orbital coupling (SOC), the band structures with SOC will be discussed next, considering the existence of transition metal M in the MSi$_2$N$_4$ MLs. The band structures with SOC with magnetic axis along [001] are calculated. The calculation results are shown in the Fig S7a-d. For CrSi$_2$N$_4$, MnSi$_2$N$_4$ and FeSi$_2$N$_4$ MLs, the bands cross the Fermi-level. Therefore, they are metals. More specifically, MnSi$_2$N$_4$ ML is a semiconductor based on the SOC calculation results, while it is a metal based on the band structures without SOC. For CoSi$_2$N$_4$ ML, the VBM of it is located at the Γ point, while the CBM is located at the F point, shown in Fig S7a-d. Therefore, it is a semiconductor with an indirect band of 0.370 eV. More details could be found in the Supporting Information.

**3.3. Electronic and Magnetic Properties of Supercell.** The 2×2×1 cells (supercell) of MSi$_2$N$_4$ MLs are also investigated to study electromagnetic properties of MSi$_2$N$_4$ family. The M atoms' spin is researched to ascertain magnetic order, as shown in Fig 4a-c. Every M atom contributes 2.51 (Cr), 3.31 (Mn), 3.72 (Fe), and 2.55 (Co) $\mu_B$ MM and there are four M atoms in a supercell. For the FM order of MSi$_2$N$_4$ MLs, there are 8.00 (CrSi$_2$N$_4$), 12.00 (MnSi$_2$N$_4$), 16.00 (FeSi$_2$N$_4$), and 11.35 (CoSi$_2$N$_4$) $\mu_B$ MM, respectively. So, the MM of MSi$_2$N$_4$ is mainly contributed by M atoms, as shown in Table 2. Besides FM order, the AFM configurations named AFM-stripy (AFM-S) is considered in research. For the MSi$_2$N$_4$ MLs' AFM orders, two M atoms contribute 2.173, 3.194, 3.640, and 2.536 $\mu_B$ while the other two M atoms contribute -2.172, -3.194, -3.640, and -2.536 $\mu_B$, respectively. Therefore, the total MM of the supercell is (closed to) 0.0 $\mu_B$ for all the MSi$_2$N$_4$ MLs. The spin charge density of the MSi$_2$N$_4$ with FM and AFM-S orders are shown in Fig 4a-c, respectively. MSi$_2$N$_4$, we define $\Delta E$ as the energy differences between different magnetic orders of MSi$_2$N$_4$ MLs. The energies of AFM-S orders of MSi$_2$N$_4$ MLs are set as 0 eV. The result shows that the energies of the FM orders are the highest among all calculated orders of the MSi$_2$N$_4$ MLs. The corresponding energy differences are shown in Table 2.

**Table 2.** MM ($\mu_B$) and the System Energy Differences $\Delta E_{FM-AFM}$ (meV) of MSi$_2$N$_4$ MLs between FM and AFM-S Orders. The MM/M appeared in the table represents the moment contributed by M atoms

| System | $\Delta E_{FM-AFM}$ | MM | | MM/M | |
|---|---|---|---|---|---|
| | | FM | AFM | FM | AFM |
| CrSi$_2$N$_4$ | 438 | 8.00 | 0 | 2.51 | ± 2.17 |
| MnSi$_2$N$_4$ | 1235 | 12.00 | 0 | 3.31 | ± 3.19 |
| FeSi$_2$N$_4$ | 547 | 16.00 | 0 | 3.72 | ± 3.64 |
| CoSi$_2$N$_4$ | 138 | 11.35 | 0 | 2.55 | ± 2.54 |

Why MSi$_2$N$_4$ MLs intend the AFM orders? The corresponding mechanism has been investigated. Each M atom is coordinated by six ligands-N in MSi$_2$N$_4$ MLs. The N-Cr-N bond angles are 69.898°, 90.440°, and 131.615°, and the N-Mn-N bond angles are 69.539°, 90.703°, and

131.498°. For FeSi$_2$N$_4$ ML, there N-Fe-N bond angles are 69.539°, 90.702°, and 131.498°. The N-Co-N bond angles are 70.113°, 90.294°, and 131.679°, respectively. In addition, the IDOS of d electrons shows that the M atoms have 4-7 electrons in d orbitals. There is direct exchange interaction, resulting in AFM coupling. Meanwhile, there is FM coupling which results from superexchange interactions according to Goodnough-Kanamori-Anderson rule[54-56] of the superexchange theorem. There are two competing exchange interactions, and the direct exchange interaction between the M atoms plays a more important role in determining the ground state, as shown in Fig S8a. In other words, the direct exchange interaction between the M atoms is stronger than superexchange interaction, as shown in Fig S8a,b. Based on the above presentation, the MSi$_2$N$_4$ show an AFM ground state unlike MoSi$_2$N$_4$.[28] The MSi$_2$N$_4$ MLs with FM orders are HM (CrSi$_2$N$_4$ and CoSi$_2$N$_4$) or semiconductors (MnSi$_2$N$_4$ and FeSi$_2$N$_4$). The MSi$_2$N$_4$ MLs with AFM-S orders are spin-unpolarized metals (CrSi$_2$N$_4$ and CoSi$_2$N$_4$) or semiconductors (MnSi$_2$N$_4$ and FeSi$_2$N$_4$), as shown in Fig 4d-g, whose gaps are 1.539 and 0.234 eV, respectively. More detail about the magnetic and electronic properties of the MSi$_2$N$_4$ MLs are presented in Fig S9 in the supporting information.

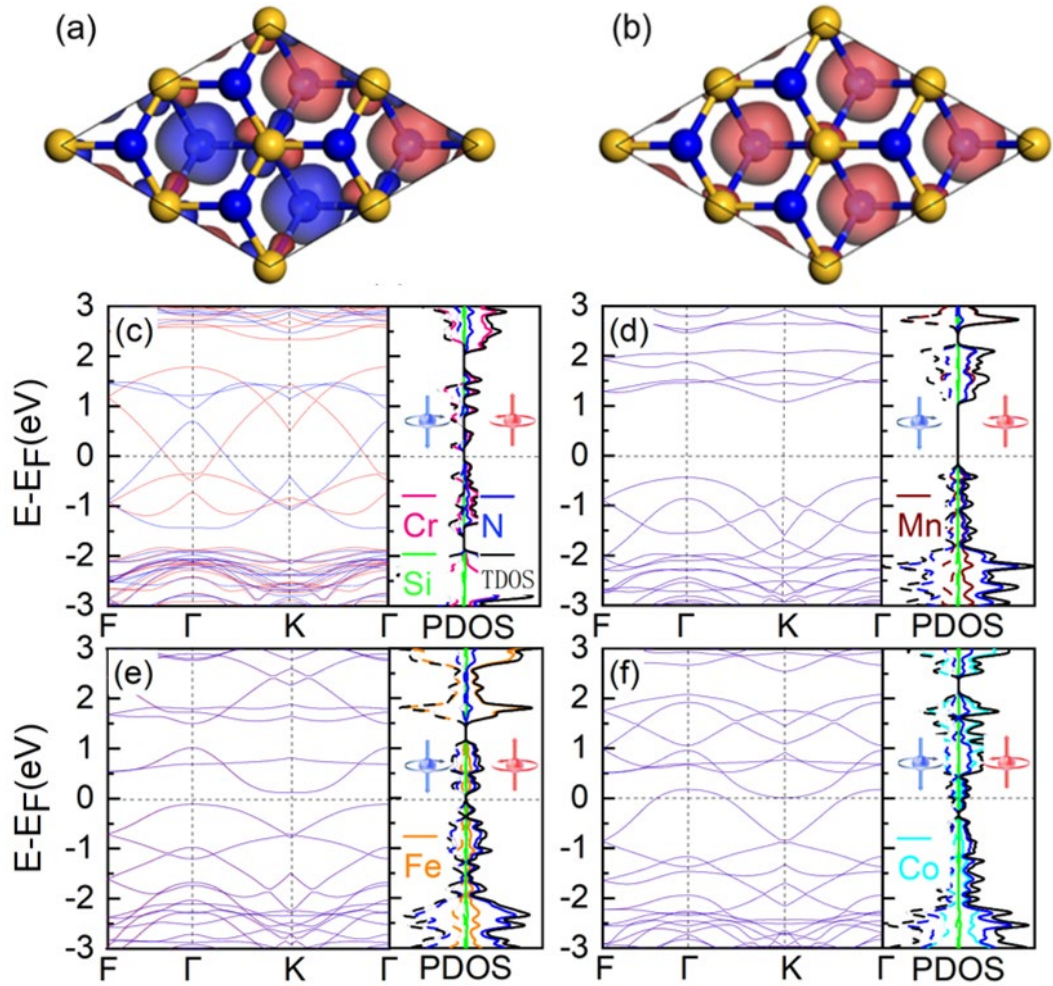

**Fig 4.** Spin charge densities of (a) AFM-S and (b) FM orders of $CoSi_2N_4$. The isovalue is 0.05 $e/Å^3$ in (a) and 0.04 $e/Å^3$ in (b). The red and blue colors represent spin-α and spin-β electrons, respectively. The spin-polarized band structure and PDOS of (c) $CrSi_2N_4$, (d) $MnSi_2N_4$, (e) $FeSi_2N_4$, and (f) $CoSi_2N_4$ MLs, respectively. The solid and dashed lines represent PDOS of spin-α and spin-β electrons, respectively.

**3.4. Electronic and Magnetic Properties of Doped $MSi_2N_4$.** Typically, the charges may be introduced during the synthesis of 2D ML materials,[57] either positive or negative. Besides, charge doping method[42, 57] is often used in tuning the relevant performance of 2D materials. Given the above, further research is required to explore the impact of carrier doping on MSi$_2$N$_4$ MLs' magnetic and electronic properties. The calculated results show that charge doping has an influence on the properties of MSi$_2$N$_4$ MLs. The carrier doping in the MSi$_2$N$_4$ MLs can lead to the magnetic phase transition as the number and structure of valence electrons of MSi$_2$N$_4$ are changed when the charges are doped, and this change affect the energy of each magnetic configuration based on Goodenough−Kanamori−Anderson rules. Besides, it may also introduce the electronic phase transition, transforming MSi$_2$N$_4$ MLs from semiconductor to HM. It is important to realize magnetic configuration regulation of MSi$_2$N$_4$ as 2D ferromagnetic materials have broad prospects in integrated chips, spintronic devices, magnetic storage, quantum information and so on.[58]

The charge doping could cause magnetic phase transformation from AFM to FM orders in MSi$_2$N$_4$. The energy differences between AFM-S orders and FM orders ($\Delta E$) are defined as $\Delta E = E_{\text{AFM}} - E_{\text{FM}}$, where $E_{\text{AFM}}$ means the energy of the MSi$_2$N$_4$ with AFM order, while $E_{\text{FM}}$ means that with FM order. The specific relationship between $\Delta E$ and doped charges is shown in Fig 5a-d. For CrSi$_2$N$_4$ ML, when -0.3 $e$ charges are doped, the $\Delta E$ is -0.816 eV. As more negative charges are introduced, the $\Delta E$ is decreased to -0.982 eV (-0.6 $e$) and -1.110 eV (-0.9 $e$), respectively. When the holes (positive charges) are introduced, the $\Delta E$ is increased to -0.442 eV (+0.1 $e$), 0.194 (+0.3 $e$), 1.078 (+0.6 $e$), and 1.646 eV (+0.9 $e$), respectively. The magnetic phase transition is observed in the CrSi$_2$N$_4$ ML when around +0.25 $e$ holes are doped, as shown in Fig 5a. ΔE increases as more positive charges or less negative charge are doped. It means that the superexchange interaction is strengthened and the direct exchange interaction is weakened relatively.

For MnSi$_2$N$_4$, $\Delta E$ firstly decreases, and then increases. As shown in Fig 5b, the energy difference between FM and AFM orders ($\Delta E$) reaches minimum when the MnSi$_2$N$_4$ ML is undoped. The minimum value of the ΔE curve is -1.235 eV. The magnetic phase transition is observed in the MnSi$_2$N$_4$ ML when around +0.54 $e$ holes are introduced, as shown in Fig 5b. The energy difference curve of FeSi2N4 is similar to that of MnSi$_2$N$_4$. Both of the $\Delta E$ curves of FeSi$_2$N$_4$ and MnSi$_2$N$_4$ MLs are V-shaped, as shown in Fig 5b,c. FeSi$_2$N$_4$ always presents an AFM order, while CoSi$_2$N$_4$ with carrier doping presents an FM order in most cases, as shown in Fig 5d.

Carrier doping could also change electronic properties. As shown in Fig 6, the undoped CrSi$_2$N$_4$ is a metal, and the undoped MnSi$_2$N$_4$ a semiconductor, respectively. When, -0.8 or -0.3 $e$ electrons are doped in CrSi$_2$N$_4$, CrSi$_2$N$_4$ is a metal. If +0.2 $e$ or more holes are introduced, CrSi$_2$N$_4$

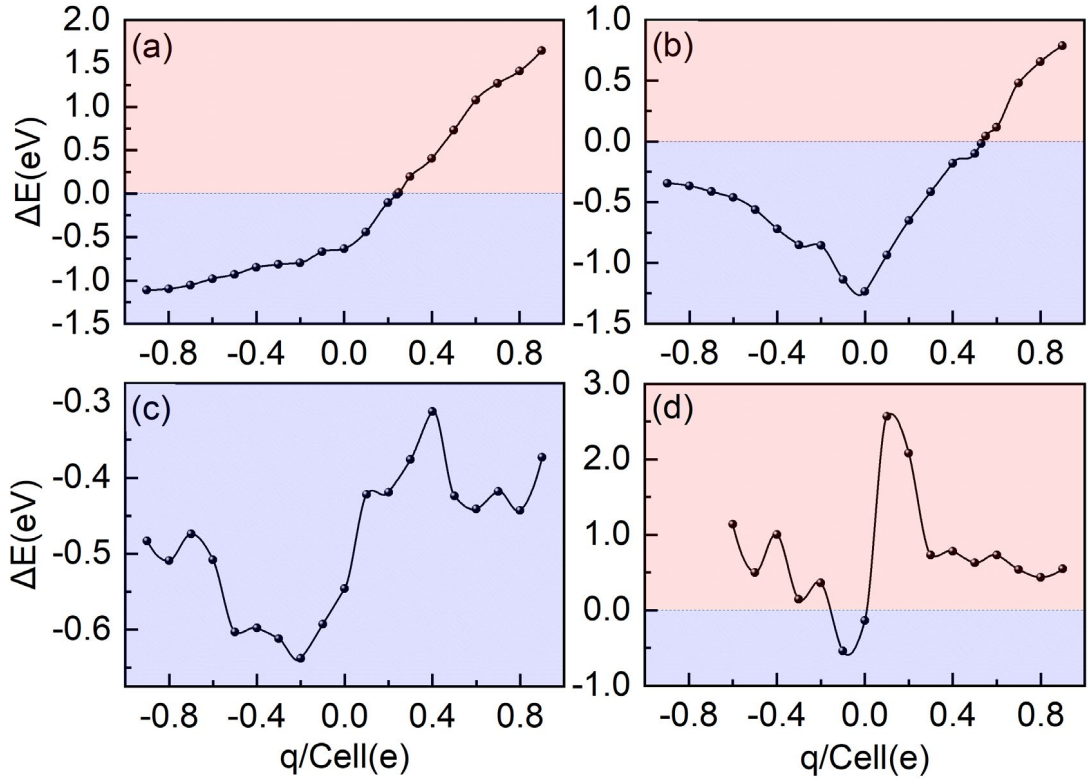

**Fig 5.** Electron and hole dope the MSi$_2$N$_4$ MLs. ΔE of (a) CrSi$_2$N$_4$, (b) MnSi$_2$N$_4$, (c) FeSi$_2$N$_4$, (d) CoSi$_2$N$_4$ change with doped charges. The q in this figure presents the number of doped charges.

ML is changed into HM, as shown in Fig 6c. It is worth noting that when +0.2 *e* hole is introduced, this change happens, and it indicates that the less hole is required for this change compared to the transformation of the magnetic orders. The states near the Fermi-level of the HM are mainly contributed by spin-α electrons, as shown in Fig 6c-e. So, why does a transition from semiconductor to HM occurs? Some electrons originally occupying the Fermi-level will be "pumped" out of the VBM of CrSi$_2$N$_4$ when positive charges are introduced. This carrier doping changes the Fermi surface and then result in the transition. For MnSi$_2$N$_4$, when the -0.6 or -0.3 *e* charges are doped, MnSi$_2$N$_4$ is a metal, while it is a semiconductor without doping. If +0.6 or +0.9 $e$ electrons are introduced, MnSi$_2$N$_4$ is converted to HM, as shown in Fig 6f,g. The states at the MnSi$_2$N$_4$ ML's Fermi- level of the HM are mainly contributed by the spin-$\alpha$ electrons like CrSi$_2$N$_4$. In addition, HM behavior is maintained when the charges are introduced in CoSi$_2$N$_4$ ML and the detail of it is shown in Fig S10 of the supporting information.

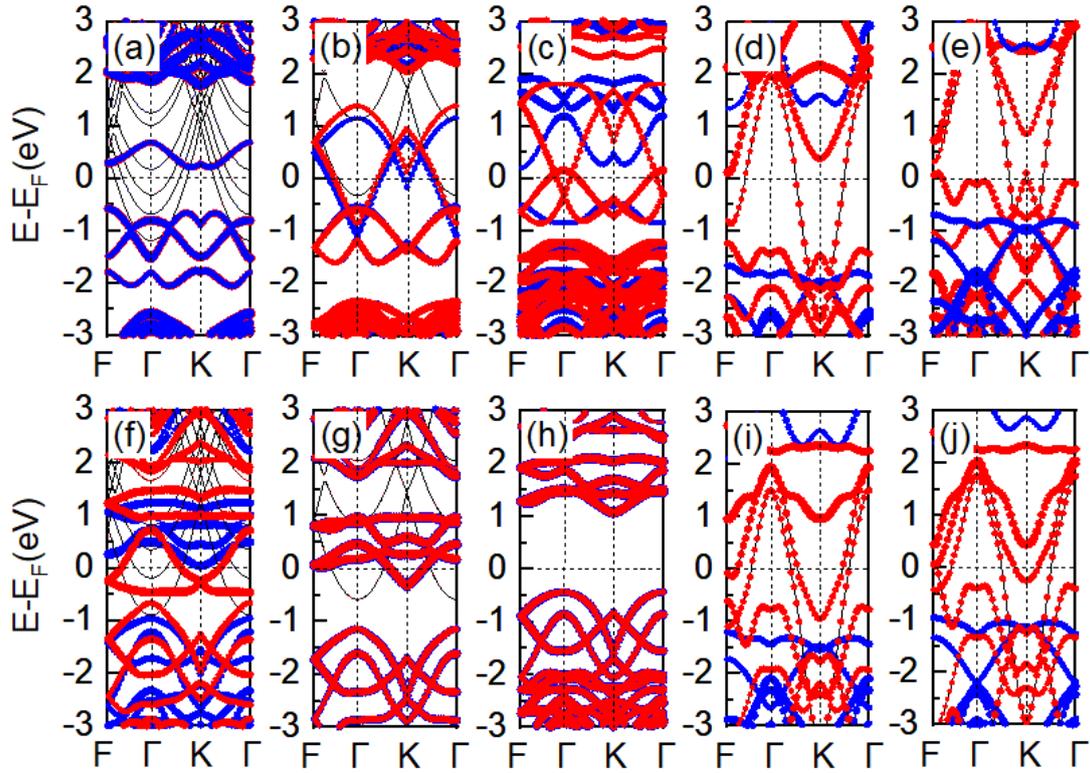

**Fig 6.** Band structures of the electrons or holes doped CrSi$_2$N$_4$ and MnSi$_2$N$_4$ MLs, calculated with HSE06 functional. The band structures of CrSi$_2$N$_4$ are doped with (a) -0.8 (b) -0.3, (c) 0.2, (d) 0.5, (e) +0.8 $e$, and that of MnSi$_2$N$_4$ are introduced with (f) -0.6, (g) -0.3, (h) 0, (i) 0.6, (j) 0.9 $e$, respectively. The red and blue lines represent spin-$\alpha$ and spin-$\beta$ electrons, respectively.

**3.5. Magnetocrystalline Anisotropy.** We usually define the energy as MAE, which is to overcome a barrier. The barrier originates from changing the magnetic moment of magnetic materials from EA to hard axis.[35, 59] The elevation of the magnetic axis is defined as $\theta$ and the azimuth of the magnetic axis is defined as $\phi$. The relation between the energies of the systems with high symmetry and $(\theta, \phi)$ is as follows:[60]

$$\begin{aligned}\Delta E_0 &= K_1\cos^2\theta + K_2\cos^4\theta + K_3\cos 3\phi \\ &= E - E_{[001]}\end{aligned} \quad (1)$$

In the above equation, $E$ is the energy in a certain elevation and azimuth $(\theta, \phi)$. $E_{001}$ represents the energy of the MSi$_2$N$_4$ MLs whose EA is along [001] direction. Similarly, $E_{[100]}$ and $E_{[010]}$ could be defined. $K_1$ and $K_2$ are the quadratic and quadratic coefficients in the eq 1, respectively. The relationship between $\Delta E_0$ and $(\theta, \phi)$ is shown in Fig 7a. Previous literature[61] indicates that $\Delta E_0$ and $\phi$ are independent. Besides, the conclusion could be drawn from the symmetry of the MSi$_2$N$_4$ MLs that each of the MSi$_2$N$_4$ MLs has a $D_{3h}$ point group. Above all, $K_3$ is equal to 0. Therefore, the eq 1 is corrected into the follow equation:

$$\Delta E_0 = K_1\cos^2\theta + K_2\cos^4\theta \quad (2)$$

For MnSi$_2$N$_4$ ML, when $E = E_{001}$ and $\theta = \pi/2$, $\Delta E_0$ reaches the maximum value (0 meV, taken as the reference energy). However, it reaches the minimum value when $\theta = 0$. The EA tends to be IMA (the

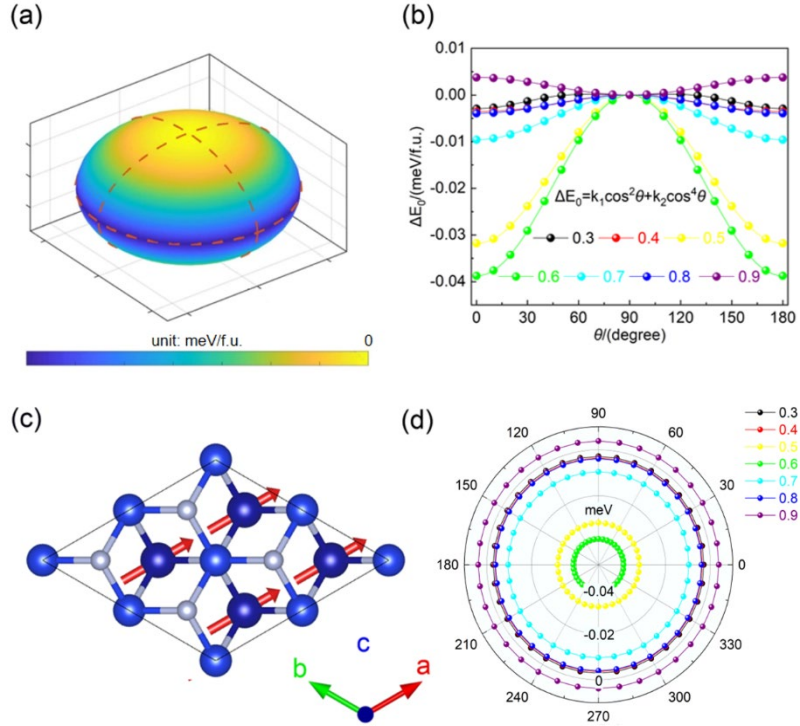

**Fig 7.** Information about MAE of CrSi$_2$N$_4$. (a) Relationship between $\Delta E_0$ and ($\theta$, $\phi$) is shown in a stereogram. (b) $\Delta E_0$ of CrSi$_2$N$_4$ with different doped charges varies as elevation changes. The black, red, yellow, green, cyan, blue and purple dotted lines represent $\Delta E_0$ of CrSi$_2$N$_4$ with doped charges of +0.3, +0.4, +0.5, +0.6, +0.7, +0.8, and +0.9 $e$ holes. (c) Red arrows represent the direction of EA of MnSi$_2$N$_4$. (d) Relationships between $\Delta E_0$ of CrSi$_2$N$_4$ and azimuthal. The black, red, yellow, green, cyan, blue and purple dotted lines represent $\Delta E_0$ of CrSi$_2$N$_4$ with doped charges of +0.3, +0.4, +0.5, +0.6, +0.7, +0.8, and +0.9 $e$.

[100] direction). The relationship between $\Delta E_0$ and $\theta$ is revealed in Fig 9a. In calculation, $U_{\text{eff}}$ is set as 4.6 eV. The relations of $\Delta E_0$ (unit: meV) of different charges doped MnSi$_2$N$_4$ ML with $\theta$ are as follows:

$$\Delta E_0 = -0.209\cos^2\theta \ (+0.6\ e)\ , \quad \Delta E_0 = -0.168\cos^2\theta \ (+0.7\ e)\ ,$$

$\Delta E_0 = -0.144\cos^2\theta \ (+0.8\ e)$, and $\Delta E_0 = -0.120\cos^2\theta \ (+0.9\ e)$. $E_{[100]}$ and $E_{[001]}$ are used to present energy of system whose $\theta$ of the magnetic axis is 0 and 90 degrees, respectively. Then MAE could be presented as the following equation:

$$\mathrm{MAE} = E_{[100]} - E_{[001]} \qquad (3)$$

When MAE is negative, the direction of the EA tends to be in the plane, and it is called as the in-plane magnetic anisotropy (IMA). By the way, when the EA prefers to be perpendicular to the plane, it is called as perpendicular magnetic anisotropy (PMA). $MnSi_2N_4$ is a 2D material, and it is expected that $MnSi_2N_4$ ML has a considerable MAE compared with three-dimensional material. The calculation results reveal that the MAE of $MnSi_2N_4$ ML is -0.209 (+0.6 $e$), -0.168 (+0.7 $e$), -0.144 (+0.8 $e$), and -0.120 meV (+0.9 $e$), respectively. Negative MAE indicates that $MnSi_2N_4$ prefers IMA, as shown in Fig 7c. The MAE of $MnSi_2N_4$ decreases with the increase of doped charge, as shown in Fig 9a,b.

The analysis and discussion of $CrSi_2N_4$ is similar to that of $MnSi_2N_4$ ML. The $U_{eff}$ is also set as 4.0 eV for $MnSi_2N_4$ ML. The relationship between $\Delta E_0$ and $\theta$ is revealed in Fig 7b. The relations between $\Delta E_0$ (unit: μeV) of $CrSi_2N_4$ ML with different carrier doping and $\theta$ are as follows:

$$\Delta E_0 = -7.45\cos^2\theta + 4.52\cos^4\theta \ (+0.3\ e) \qquad (4)$$

$$\Delta E_0 = -2.61\cos^2\theta - 0.95\cos^4\theta \ (+0.4\ e) \tag{5}$$

$$\Delta E_0 = -31.77\cos^2\theta \ (+0.5\ e) \tag{6}$$

$$\Delta E_0 = -38.76\cos^2\theta \ (+0.6\ e) \tag{7}$$

$$\Delta E_0 = -9.59\cos^2\theta \ (+0.7\ e) \tag{8}$$

$$\Delta E_0 = -3.93\cos^2\theta \ (+0.8\ e) \tag{9}$$

$$\Delta E_0 = 3.77\cos^2\theta \ (+0.9\ e) \tag{10}$$

The MAE of $CrSi_2N_4$ ML shown in Fig 7d is -2.93 (+0.3 $e$), -3.56 (+0.4 $e$), -31.78 (+0.5 $e$), -38.77 (+0.6 $e$), -9.59 (+0.7 $e$), -3.93 (+0.8 $e$), and 3.76 (+0.9 $e$) μeV, respectively. MAE of $CrSi_2N_4$ decreases first and then increases. Interestingly, when +0.9 $e$ electrons are introduced, the negative MAE value substituted for positive one, which means EA of $CrSi_2N_4$ ML has changed from [100] to [001] direction. Accordingly, there is a switch of the magnetic anisotropy of $CrSi_2N_4$ from the IMA to PMA.

It is necessary to determine the contribution of each atomic orbital to the MAE of $CrSi_2N_4$ and $MnSi_2N_4$. Besides, explaining the relationship between the MAEs (MAE of $CrSi_2N_4$ and $MnSi_2N_4$) and doped charges is necessary. To accomplish the above objectives, tight binding and second-order perturbation theory are used to calculate the MAE. According to the canonical formula,[62] each atom's contribution to MAE can be calculated by using the following formula:

**Table 3.** Symbols Appearing in the Discussion of Magnetocrystalline Anisotropy and Their Meanings

| Symbol | Meaning | Symbol | Meaning |
|---|---|---|---|
| $MAE_i$ | MAE contributed by atom i | $L_x$ | angular momentum operators in [100] direction |
| $n_i^{[100]}$ | density of state in [100] direction | $L_z$ | angular momentum operators in [001] direction |
| $n_i^{[001]}$ | density of state in [001] direction | $u$ | unoccupied state |
| $MAE_{tot}$ | total MAE | $o$ | occupied state |
| + | spin-α state | $E_u$ | energy of unoccupied state |
| - | spin-β state | $E_o$ | energy of occupied state |
| $\xi$ | SOC constant | | |

$$\mathrm{MAE}_i = \left[ \int E_f (E - E_F)[n_i^{[100]}(E) - n_i^{[001]}(E)] \right] \quad (11)$$

where the meanings of the symbols in the eq 11 are shown in Table 3. Given the symmetry of MSi$_2$N$_4$ MLs ($D_{3h}$ group), the $E_{[010]}$ is equal to

$E_{[100]}$,[33] thus the system whose EA follows [010] direction isn't considered. Besides, the following formula is true:

$$\mathrm{MAE}_{tot} = \sum_i \mathrm{MAE}_i \tag{12}$$

where the meanings of the symbols in the eq 12 are also shown in Table 3. Given the second-order perturbation theory,[63] MAE can be calculated by the following equation:

$$\Delta E^{--} = E_x^{--} - E_z^{--} = \xi^2 \sum_{o^+, u^-} (|\langle o^- | L_z | u^- \rangle|^2 - |\langle o^- | L_x | u^- \rangle|^2)/(E_u^- - E_o^-) \tag{13}$$

$$\Delta E^{-+} = E_x^{+-} - E_z^{+-} = \xi^2 \sum_{o^+, u^-} (|\langle o^+ | L_z | u^- \rangle|^2 - |\langle o^+ | L_x | u^- \rangle|^2)/(E_u^- - E_o^-) \tag{14}$$

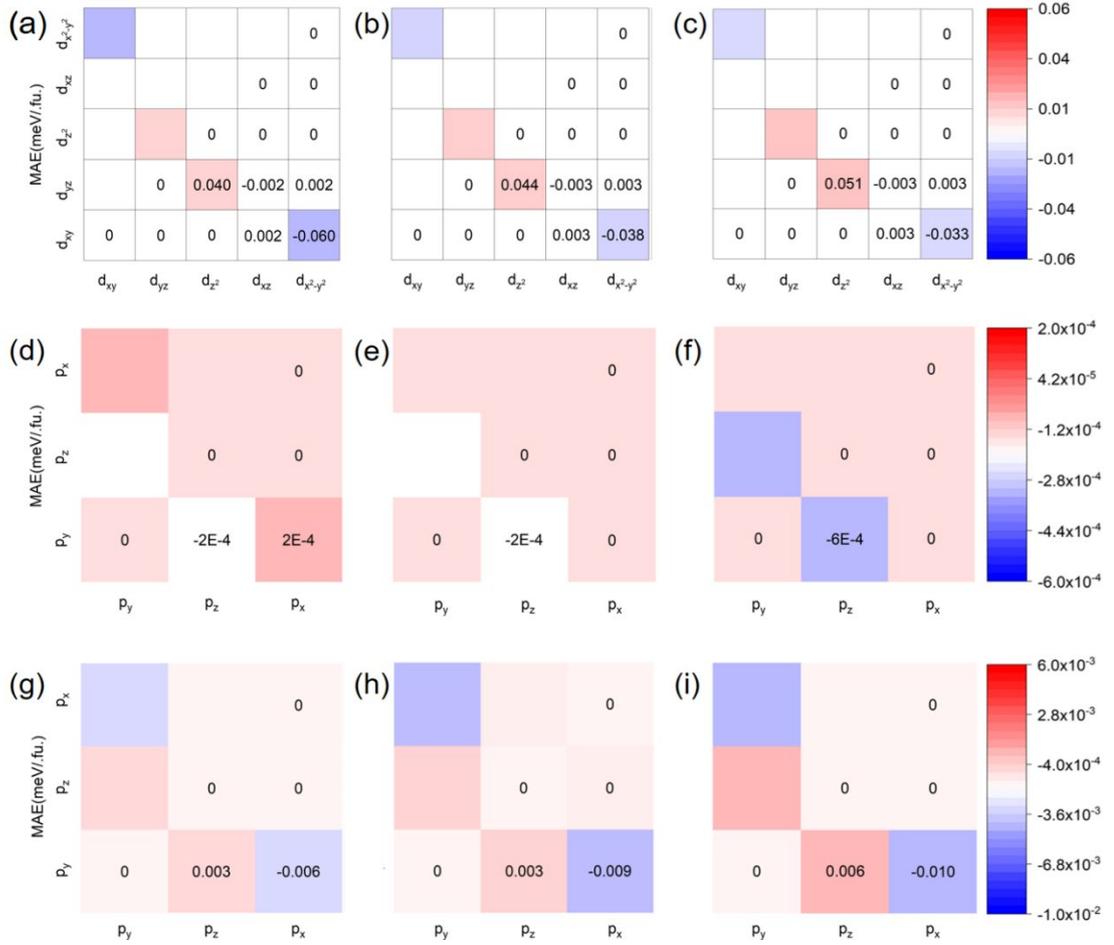

**Fig 8.** Orbital-resolved MAE of CrSi$_2$N$_4$ ML with doped charges of (a, d,

g) +0.6, (b, e, h) +0.7, and (c, f, i) +0.9 $e$, respectively. The orbital-resolved MAEs contributed by (a, b, c) Cr, (d, e, f) Si, and (g, h, i) N atoms are shown in this figure, respectively.

where the meanings of the symbols in the equation are also shown in Table 3. MAE is primarily composed of spin orbit matrix element and energy difference. According to eq 4, MAE is related to density of state near Fermi-level. The matrix element differences $|<o^-|L_z|u^->|^2 - |<o^-|L_x|u^->|^2$ and $|<o^+|L_z|u^->|^2 - |<o^+|L_x|u^->|^2$ of p and d orbitals are calculated which are shown in Table 4 and Table 5.

**Table 4.** Matrix Differences of d Orbitals along [001] and [100] Directions in Eqs 6 and 7

| $u^-$ | $o^+$ | | | | | $o^-$ | | | | |
|---|---|---|---|---|---|---|---|---|---|---|
| | $d_{xy}$ | $d_{yz}$ | $d_{z^2}$ | $d_{xz}$ | $d_{x^2-y^2}$ | $d_{xy}$ | $d_{yz}$ | $d_{z^2}$ | $d_{xz}$ | $d_{x^2-y^2}$ |
| $d_{xy}$ | 0 | 0 | 0 | 1 | -4 | 0 | 0 | 0 | -1 | 4 |
| $d_{yz}$ | 0 | 0 | 3 | -1 | 1 | 0 | 0 | -3 | 1 | -1 |
| $d_{z^2}$ | 0 | 3 | 0 | 0 | 0 | 0 | -3 | 0 | 0 | 0 |
| $d_{xz}$ | 1 | -1 | 0 | 0 | 0 | -1 | 1 | 0 | 0 | 0 |
| $d_{x^2-y^2}$ | -4 | 1 | 0 | 0 | 0 | 4 | -1 | 0 | 0 | 0 |

For furtherly explaining the change of MAE with doped charges, atomic orbital resolved MAE is researched and the corresponding results are shown in Fig 8a-i and 9c-k. For $CrSi_2N_4$, the results show that the MAE is relatively low. Each atom's contribution to MAE is shown in Table S1. The hybridization between Cr's d orbitals is also investigated. It could be concluded that the hybridization between Cr atoms' $d_{z^2}$ and $d_{yz}$ contributes positively to MAE (corresponds to the matrix differences 3 for the d-orbitals) from 0.040 (+0.6 $e$) to 0.051 (+0.9 $e$) meV. The hybridization between Cr atoms' $d_{x^2-y^2}$ and $d_{xy}$ contributes negatively to MAE (corresponds to the matrix differences -4 for the Cr atom's d-orbitals while the contribution of Si and N atoms is different) from -0.060 (+0.6 $e$) to -0.033 (+0.9 $e$) meV. More information about the hybridization between

**Table 5.** Matrix Differences of p Orbitals along [001] and [100] Directions in Eqs 6 and 7

| | $o^+$ | | | $o^-$ | | |
|---|---|---|---|---|---|---|
| $u^-$ | $p_y$ | $p_z$ | $p_x$ | $p_y$ | $p_z$ | $p_x$ |
| $p_y$ | 0 | 1 | -1 | 0 | -1 | 1 |
| $p_z$ | 1 | 0 | 0 | -1 | 0 | 0 |
| $p_x$ | -1 | 0 | 0 | 1 | 0 | 0 |

Cr's d orbitals could be found in Fig 8a-i. According to the above result, we can conclude that the positive contribution (to MAE) of the hybridization between Cr atoms' $d_{z^2}$ and $d_{yz}$ increases with the doped positive charges. We also could conclude that the negative contribution of the hybridization between Cr atoms' $d_{x^2-y^2}$ and $d_{xy}$ decreases with the doped charges. This could explain that the MAE sign reverses. The MAE sign reverses, which means the transformation of EA of $CrSi_2N_4$.

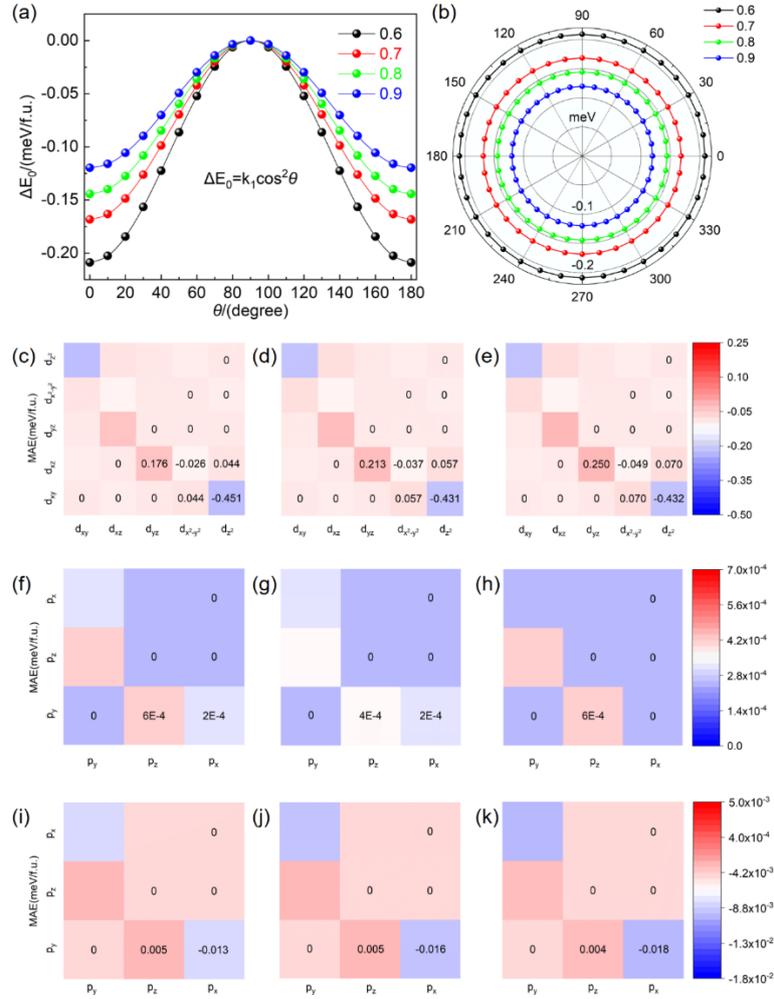

**Fig 9.** (a) $\Delta E_0$ of the doped $MnSi_2N_4$ ML varies as elevation changes. The black, red, green and blue dotted lines represent $\Delta E_0$ of $MnSi_2N_4$ with +0.6, +0.7, +0.8, and +0.9 $e$ doped charges varies as elevation changes. (b) $\Delta E_0$ of $MnSi_2N_4$ with different doped charges varies as azimuthal changes. The black, red, green and blue dotted lines represent $\Delta E_0$ of $MnSi_2N_4$ with +0.6, +0.7, +0.8, and +0.9 $e$ doped charges varies as azimuthal changes.

Orbital-resolved MAE of MnSi$_2$N$_4$ ML with doped charges of (c, f, i) +0.6, (d, g, j) +0.8, and (e, h, k) +1.0 $e$, respectively. The orbital-resolved MAEs contributed by (c, d, e) Mn, (f, g, h) Si, and (i, j, k) N atoms are shown in Fig 9, respectively.

**3.6. Dynamical and Thermal Stability.** The phonon dispersion curve and the PHDOS are calculated, as shown in Fig 10a-d. The results show there is no imaginary phonon mode, thus the dynamic stability of MSi$_2$N$_4$ is conformed. For the PHDOS contributed by M, the highest vibration frequency is 11.97 (Cr), 6.53 (Mn), 5.28 (Fe) and 5.23 (Co) THz, respectively, which decreases with atomic number increasing. In addition, it can be found that the PHDOS is mainly contributed by M atom for the low frequency (0 < $\varepsilon$ < 10 THz) while it is mainly contributed by N and Si atoms for the high frequency ($\varepsilon$ > 10 THz). This difference is due to the different atomic weight between Si, N and M atoms. More specifically, the M atoms (atomic weight: 52.00, 54.94, 55.85, and 58.93) are heavier than Si (atomic weight: 28.09) and N atoms (atomic weight: 14.01). Therefore, the high frequency phonons are mainly contributed by the silicon and nitrogen atoms, while the low frequency phonons are mainly contributed by the M atoms, as shown in Fig 10a-d. M atoms contribute thermal conductivity majorly at low temperatures by the low-frequency phonons, however, thermal conductivity is majorly contributed by silicon and nitrogen atoms at high temperatures with the high-frequency phonons.

Since the MSi$_2$N$_4$ MLs are composed of the same non-metallic atoms, and these non-metallic atoms is lighter than the M and Mo atoms, the maximum vibration frequency of MSi$_2$N$_4$ is close to that of MoSi$_2$N$_4$.[64]

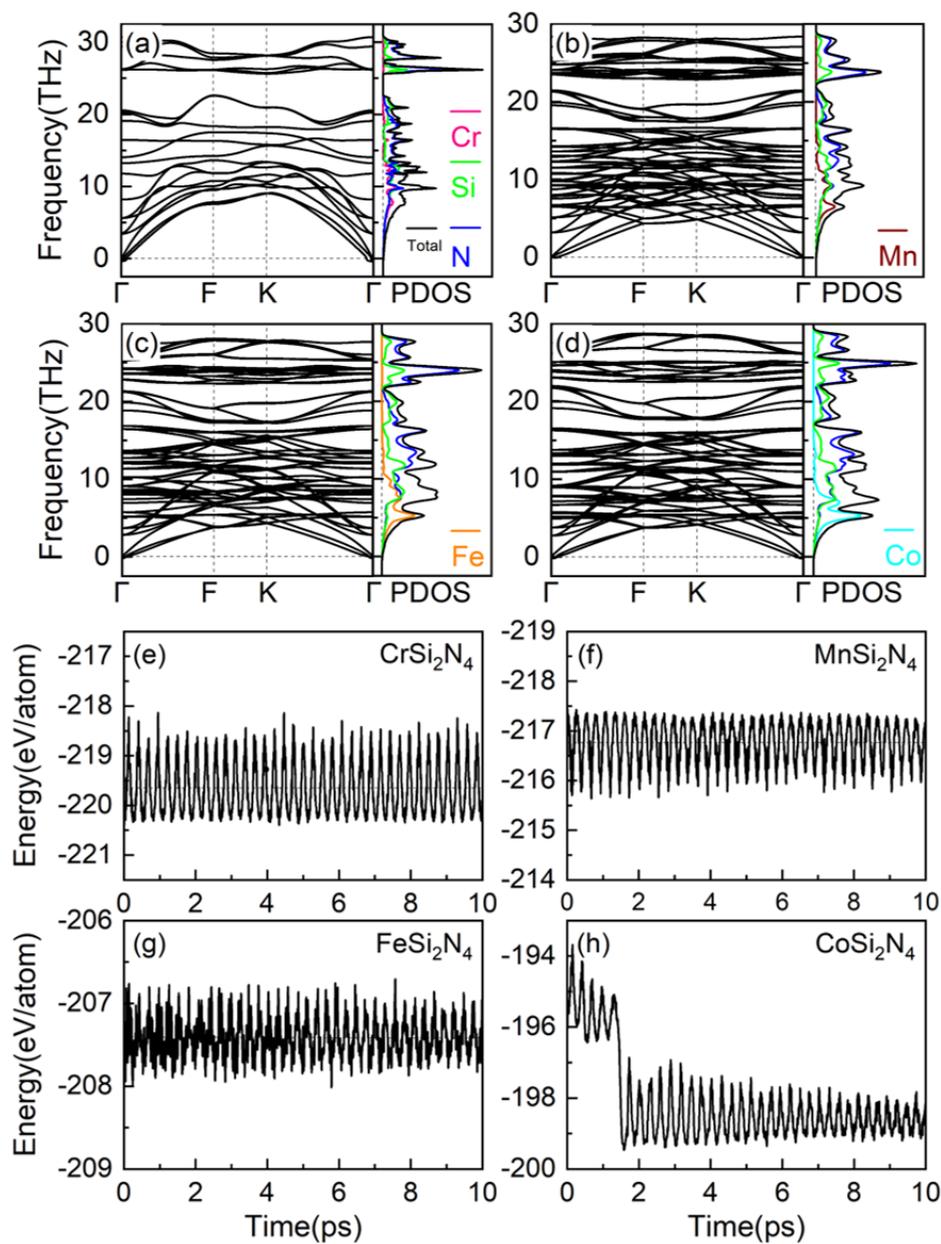

**Fig 10.** (a−d) Phonon spectrum and PHDOS of (a) CrSi$_2$N$_4$, (b) MnSi$_2$N$_4$, (c) FeSi$_2$N$_4$, and (d) CoSi$_2$N$_4$ MLs. The green, blue, pink, wine red, orange, cyan lines represent Si, N, Cr, Mn, Fe, and Co atoms' projected phonon DOS, respectively. (e−h) AIMD of evaluation of energies at the PBE + U

level for 10 ps at 300 K. The total energies of (e) $CrSi_2N_4$, (f) $MnSi_2N_4$, (g) $FeSi_2N_4$, and (h) $CoSi_2N_4$ are shown in Fig 10e-h, respectively.

To confirm the thermal stability of $MSi_2N_4$ family, the widely-used AIMD simulation which is applied in studying the stability of materials is performed at 300K. The result shows that the total energies of $MSi_2N_4$ fluctuated around -219.65 (Cr), -216.77 (Mn), -207.41 (Fe), and -198.58 eV (Co) at 300 K, as shown in Fig 10 e-h. The total energies of $MSi_2N_4$ increase with the atomic number. The phase transition is occurred in $CoSi_2N_4$ ML, and the energy of $CoSi_2N_4$ ML decreases during this phase transition. More information about the dynamical and thermal stability of the $MSi_2N_4$ is shown in Fig S11 in the supporting information.

## 4. CONCLUSIONS

In a word, DFT calculation is used to study the structure, electronic and magnetic properties of $MSi_2N_4$ MLs. We analyze the electromagnetic properties of intrinsic $MSi_2N_4$ MLs, and find that the ground states of them are metals or semiconductors with the AFM orders. $MSi_2N_4$ MLs' AFM orders mainly result from their direct exchange interactions. We have studied the effect of carrier doping on $MSi_2N_4$ MLs. The results show that charge doping will transform the electronic properties (mainly from semiconductor to HM) and magnetic phase of the $MSi_2N_4$ MLs. By using noncollinear DFT calculations, the magnetocrystalline anisotropy of the doped $MSi_2N_4$ ML with FM order is estimated. For the doped $CrSi_2N_4$ ML with FM order, the carrier doping could transform the EA from [100] direction to [001] direction. The results show that $MSi_2N_4$ MLs have excellent tunable electromagnetic properties. We believe that our research will provide an impetus for further theoretical and experimental research in the field of 2D magnetic materials.

**Author Information**

Corresponding Author

*E-mail: suya@sdu.edu.cn; Tel: +86-0531-88392369; Fax: +86-0531-88392369

*E-mail: zyguan@sdu.edu.cn; Tel: +86-0531-88363179; Fax: +86-0531-88363179

**Acknowledgements**

This work was supported by the financial support from the Natural Science Foundation of China (Grant No. 11904203), and the Fundamental Research Funds of Shandong University (Grant No. 2019GN065). The computational resources from Shanghai Supercomputer Center. The scientific calculation in this paper have been performed on the HPC Cloud Platform of Shandong University. The authors are grateful to Beijing Beilong Super Cloud Computing Co., Ltd for the computation resource in the Beijing Super Cloud Computing Center. The authors are grateful to Tencent Quantum Laboratory for the computation resource.

**Conflict of Interest:** The authors declare no competing financial interest.